\documentclass[final]{mn2e}
\usepackage{umlaut}
\usepackage{color}
\usepackage[sectionbib]{natbib}
\usepackage{chapterbib}
\usepackage{graphicx}
\usepackage{graphics}
\usepackage{rotating}
\usepackage{amsmath}        
\usepackage{amssymb}

\date{This is a preprint of an Article accepted for publication in
  MNRAS.\\ \copyright\ 2006 RAS}
\pagerange{\pageref{firstpage}--\pageref{lastpage}}
\pubyear{2006}

\def\LaTeX{L\kern-.36em\raise.3ex\hbox{a}\kern-.15em
    T\kern-.1667em\lower.7ex\hbox{E}\kern-.125emX}

\begin{document}

	\textheight 23cm

\newcommand{\mnras}{MNRAS}
\newcommand{\apjl}{ApJ}
\newcommand{\apj}{ApJ}
\newcommand{\aj}{AJ}
\newcommand{\nat}{Nature}
\newcommand{\aap}{A\&A}
\newcommand{\iaucirc}{IAU Circular}
\newcommand{\pasj}{PASJ}
\newcommand{\apss}{Ap\&SS}
\newcommand{\araa}{ARA\&A}
\newcommand{\pasp}{PASP}

\newcommand{\lap}{$\stackrel{\textstyle <}{\sim} \;$}
\newcommand{\gap}{$\stackrel{\textstyle >}{\sim} \;$}
\newcommand{\std}{\normalsize}
\newcommand{\sqchi}{\hbox{$\chi^{2}$}}
\newcommand{\Msun}{$M_{\odot}$}
\newcommand{\degree}{$^{\circ}$}

\label{firstpage}

\title[Jet opening angles]{Opening angles, Lorentz factors and confinement of X-ray binary jets}

\author[J.C.A.~Miller-Jones et al.]
 {J.C.A.~Miller-Jones,$^1$\thanks{email: jmiller@science.uva.nl}
 R.P.~Fender,$^2$ and E.~Nakar.$^3$\\
$^1$Astronomical Institute 'Anton Pannekoek', University of Amsterdam, Kruislaan  403,\\
1098 SJ, Amsterdam, The Netherlands.\\
$^2$School of Physics and Astronomy, University of Southampton,
Highfield, Southampton, SO17 1BJ, UK\\
$^3$Theoretical Astrophysics, Caltech, Pasadena, CA 91125, USA}

\maketitle

\begin{abstract}
We present a collation of the available data on the opening angles of
jets in X-ray binaries, which in most cases are small
($\lesssim10$\degree).  Under the assumption of no confinement, we
calculate the Lorentz factors required to produce such small opening
angles via the transverse relativistic Doppler effect.  The derived
Lorentz factors, which are in most cases lower limits, are found to be
large, with a mean $>10$, comparable to those estimated for AGN and
much higher than the commonly-assumed values for X-ray binaries of
2--5.  Jet power constraints do not in most cases rule out such high
Lorentz factors.  The upper limits on the opening angles show no
evidence for smaller Lorentz factors in the steady jets of Cygnus X-1
and GRS\,1915+105.  In those sources in which deceleration has been
observed (notably XTE\,J1550--564 and Cygnus X-3), some confinement of
the jets must be occurring, and we briefly discuss possible
confinement mechanisms.  It is however possible that all the jets
could be confined, in which case the requirement for high bulk Lorentz
factors can be relaxed.
\end{abstract}


\section{Introduction}
Proper motions of X-ray binary (XRB) jets have often been used to
place limits on the jet Lorentz factors.  \citet{Fen03} recently
argued that it was in fact impossible to do more than place a lower
limit on the Lorentz factors of the jets from two-sided jet proper
motions.  For the persistent, continuous jets observed to exist in the
low/hard X-ray states of black hole candidates, Gallo, Fender \&
Pooley (2003) found a universal correlation between the X-ray and
radio fluxes of the sources, and used the scatter about this relation
to constrain the Lorentz factors of such jets to $\lesssim 2$.
However, \citet{Hei04} argued that the scatter about such a relation
could not be used to constrain the mean Lorentz factor of the jets,
but rather only the width of the Lorentz factor distribution.  Other
arguments, such as those based on jet power requirements, must be used
to determine the absolute values of the jet Lorentz factors.

Bulk jet flow velocities close to $c$, the speed of light, have been
inferred in many XRB systems \citep[e.g.][]{Mir94,Hje95,Fen04},
whereas the transverse expansion speeds have not yet been reliably
measured.  To date, there are few reported detections of XRB jets
resolved perpendicular to the jet axis.  This places strong upper
limits on the opening angles of the jets (often less than a few
degrees; see Table \ref{tab:obs}), and hence on the transverse
expansion speeds.  While jets can in principle undergo transverse
expansion at a significant fraction of $c$, time dilation effects
associated with the bulk motion would reduce the apparent opening
angle in the observer's frame.  The magnitude of this effect would be
determined by the bulk Lorentz factor of the flow.  This raises the
possibility of using the observed opening angle of a freely-expanding
jet to constrain its Lorentz factor.  Alternatively, if the Lorentz
factors thus derived were incompatible with values deduced from
independent methods, a strong argument could be made for jet
confinement out to large (parsec) scales in such Galactic sources.

In this paper, we first develop the formalism for deducing the Lorentz
factor of a freely-expanding jet given a measurement of the opening
angle and the inclination angle of the jet axis to the line of sight.
In \S\,\ref{sec:constraints} we compare constraints on the Lorentz
factors derived from opening angle considerations with those from
other methods.  Transient and steady jets are compared in
\S\,\ref{sec:low-hard}, and the derived X-ray binary jet Lorentz
factors are compared to those seen in AGN in \S\,\ref{sec:AGN}. We
discuss possible mechanisms for jet confinement and a method of using
lightcurves to test confinement in \S\,\ref{sec:confinement}.  A
summary of the observed properties of the individual sources is given
in Appendix \ref{sec:details}.

\section{Formalism}
\label{sec:formalism}
\begin{figure}
\begin{center}
\includegraphics[width=0.45\textwidth,angle=0,clip=]{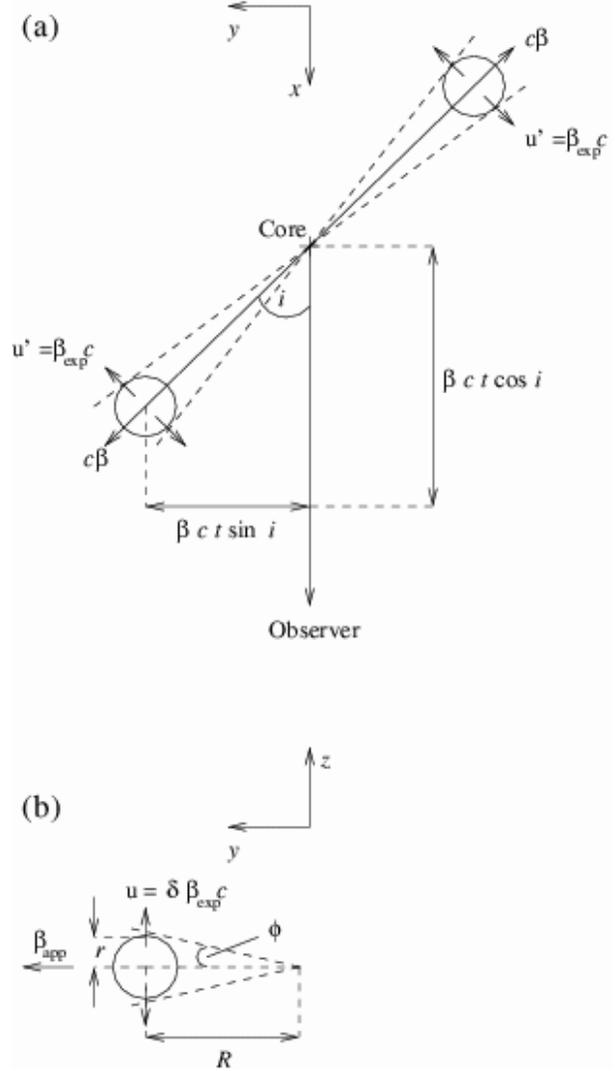}
\caption{Schematic of the geometry of the jets.  (a) shows the plane
  of the knot motion, and (b) shows the motion as seen by the
  observer.}
\label{fig:geometry}
\end{center}
\end{figure}

We consider the case of knots (although we note that they could be
internal shocks rather than plasmons) propagating in an XRB jet.  To
simplify matters, we consider a single, spherical knot expanding
radially with constant velocity $u^{\prime} = dr^{\prime}/dt^{\prime}$
in its own frame.  Primed quantities are measured in the frame of the
moving knot and unprimed quantities in the frame of the stationary
observer.  The expansion speed of the knot as seen in the observer's
frame is modified by the relativistic Doppler factor, and is thus
\begin{equation}
u = \frac{u^{\prime}}{\Gamma(1-\beta\cos i)} = \delta u^{\prime},
\end{equation}
where $\delta$ is the Doppler factor $\Gamma^{-1}(1-\beta\cos
i)^{-1}$, $\beta c$ is the jet speed, $\Gamma=(1-\beta^2)^{-1/2}$ is
the bulk Lorentz factor of the jet, and $i$ is the inclination angle
of the jet axis to the line of sight.  The geometry is shown in
Fig.\,\ref{fig:geometry}.

In most cases, we do not observe the actual expansion velocity
directly.  Most of the observations simply place a constraint on the
half-opening angle $\phi$ of the jet from its width at a given
distance from the core.  The observed knot radius at a given time $t$
(in the observer's frame) since the knot was ejected is
$r = ut$.  The apparent speed of the knot away from the core is
$\beta_{\rm app} c$, given by
\begin{equation}
\beta_{\rm app} = \frac{\beta\sin i}{1-\beta\cos i},
\label{eq:vel_apparent}
\end{equation}
having taken into account the combination of projection effects and
the motion of the knot towards the observer.  Thus, since the observed
distance of the knot from the core is $R = \beta_{\rm app}ct$, then
\begin{equation}
\tan\phi = \frac{r}{R} = \frac{u^{\prime}}{\Gamma\beta c \sin i}.
\label{eq:opening_angle}
\end{equation}
Rearranging, we can therefore derive an expression for the bulk jet
Lorentz factor implied by the measured opening angles, writing the
intrinsic jet knot expansion speed as $u^{\prime} = \beta_{\rm exp}
c$.  Thus
\begin{equation}
\Gamma = \left(1+\frac{\beta_{\rm exp}^2}{\tan^2\phi\sin^2
i}\right)^{1/2}.
\label{eq:gamma}
\end{equation}
Whereas $i$ alone is difficult to constrain, in many cases $\beta\cos
i$ is a measurable variable.  Knowing $\phi$, we can then find
$\Gamma$ as a function of $i$, from $i=0$ up to $i_{\rm max} =
\cos^{-1}(\beta\cos i)$.  This can then be compared at each value of
$i$ to the value of $\Gamma = (1-\beta^2)^{-1/2}$ obtained from
$\beta\cos i$.

$\beta_{\rm exp}$ is commonly assumed to have an upper limit
corresponding to the relativistic sound speed, $c/\sqrt{3}$.  However,
the acceleration of a cloud of hot jet plasma that expands under its
own pressure in a zero-pressure external medium without any
confinement is not done by a sound wave, so the expansion velocity is
limited by the initial thermal velocity and not by the sound velocity
\citep{Lan59}.  Thus a jet that is initially relativistically hot can
in principle undergo transverse expansion at the speed of light
\citep[e.g.][]{Sar99}.  We therefore performed our calculations up to the
limit $\beta_{\rm exp}=1$.

\section{Constraints on the Lorentz factors}
\label{sec:constraints}
\subsection{Observational sample}
\label{sec:literature}

\begin{table*}
\begin{center}
\begin{tabular}{lcccccccc} \hline \hline
System & $\phi$ & $\beta\cos i$ & $\Gamma_{\rm min,exp}$ & $d_{\rm
  max}$ (kpc) & Reference & $d$ (kpc) & Reference & $\Gamma_{\rm d}$\\
\hline
GRS\,1915+105 (steady) & $\leq 4.9$\degree & $0.035\pm0.017$ & 11.7 & & D00b &
6.1--12.2 & D00 & \\
GRS\,1915+105 (transient)& $\leq 4$\degree & $0.41\pm0.02$ & 15.7 &
  $11.2\pm0.8$ & F99 &
6.1--12.2 & D00 & $>1.8$\\
Cygnus X-3 (small-scale)& $5.0\pm0.5$\degree & $0.50\pm0.10$ & 13.3 &
$35.7\pm4.8$ & MJ04 & $\sim 10$ & D83 & $1.2\pm0.1$\\
Cygnus X-3 (large-scale) & $<16.5$\degree & $0.14\pm0.03$ & 3.6 &
$21\pm1$ & M01 & $\sim 10$ & D83 & $1.2\pm0.1$\\
GRO\,J1655--40 & $\lesssim3.1$\degree & $0.091\pm0.014$ & 18.6 & $3.5\pm0.1$
& HR95 & $3.2\pm0.2$ & HR95 & $>2.4$\\
V4641\,Sgr & $\leq 25.1$\degree & $\sim 0.4$ & 2.6 & & H00 &
$9.59^{+2.72}_{-2.19}$ & O01 &\\
LS\,5039 & $\leq 6$\degree & $0.17\pm0.05$ & 9.7 & & P02 & $2.9\pm0.3$ & R02a
&\\
XTE\,J1550--564 & $\leq 3.7$\degree & $0.61\pm0.13$ & 19.6 & $16.5\pm3.5$ & T03
& 3.2--9.8 & O02 & $1.3\pm0.2$\\
H\,1743--322 & $\leq 6$\degree & $0.23\pm0.05$ & 9.8 & $10.4\pm2.9$ & C05 &
8.5? & &\\
Cygnus X-1 (steady) & $<2$\degree & $>0.50$ & 33.0 & & S01 &
$2.1\pm0.1$ & M95 &\\
Cygnus X-1 (transient) & $<18$\degree & $>0.2$ & 3.7 & & F05 &
$2.1\pm0.1$ & M95 &\\
GX\,339--4 & $\leq 12$\degree & $>0.16\pm0.05$ & 4.9 & & G04 & $>6$ & H04\\
1RXS\,J001442.2+580201 & $\leq 1.8$\degree & $0.20\pm0.02$ & 32.5 & & R02b &\\
\hline \hline
\end{tabular}
\caption{Summary of the observations from the literature.
  $\Gamma_{\rm min,exp}$ denotes the Lorentz factors derived from the
  opening angles (Equation\,\ref{eq:gamma}) and the constraints on
  $\beta\cos i$, and $\Gamma_{\rm d}$ the Lorentz factors derived from
  the ratio $d/d_{\rm max}$ (Equation\,\ref{eq:gamma_d}).  The first
  column of references gives the papers from which the values of
  $\phi$, $\beta\cos i$ and $d_{\rm max}$ were derived.  The second
  gives the papers from which the distance estimates were taken.  D00b
  = \citet{Dha00b}; D00 = Dhawan, Goss \& Rodr\'\i guez (2000a); F99 =
  \citet{Fen99}; MJ04 = \citet{Mil04}; D83 = \citet{Dic83}; M01 =
  \citet{Mar01}; HR95 = \citet{Hje95}; H00 = \citet{Hje00}; O01 =
  \citet{Oro01}; P02 = \citet{Par02}; R02a = \citet{Rib02a}; T03 =
  \citet{Tom03}; O02 = \citet{Oro02}; C05 = \citet{Cor05}; S01 =
  \citet{Sti01}; M95 = Massey, Johnson \& Degioia-Eastwood (1995); F05
  = Fender et al. (in prep.); G04 = \citet{Gal04}; H04 =
  \citet{Hyn04}; R02b = \citet{Rib02b}}
\label{tab:obs}
\end{center}
\end{table*}

We have amassed from the literature a compilation of the known X-ray
binaries with resolved radio jets.  A description of the individual
systems is given in Appendix \ref{sec:details}.  In most cases, the
jets are unresolved perpendicular to the jet axis, giving an upper
limit to the opening angle of the jets as observed in our frame,
determined by the beam size and the angular separation from the centre
of the system.

In those sources for which the proper motions of both the approaching
and receding jet knots, $\mu_{\rm a}$ and $\mu_{\rm
r}$ respectively, can be measured, the product
\begin{equation}
\beta\cos i = \frac{\mu_{\rm a}-\mu_{\rm r}}{\mu_{\rm a}+\mu_{\rm r}},
\end{equation}
may be calculated, constraining the values of $\beta>\beta\cos i$ and
$i<\cos^{-1}(\beta\cos i)$.  For a given source distance $d$, it is
possible to solve for the exact values of $\beta$ and $i$.  There is
therefore a maximum possible distance to the source,
\begin{equation}
d_{\rm max} = \frac{c}{\sqrt{\mu_{\rm a}\mu_{\rm r}}},
\label{eq:dmax}
\end{equation}
which corresponds to the maximum possible intrinsic jet speed,
$\beta=1$.

\subsection{Lorentz factors from opening angle constraints}
\label{sec:opening_angle}
\begin{figure*}
\begin{center}
\includegraphics[width=0.9\textwidth,angle=0,clip=]{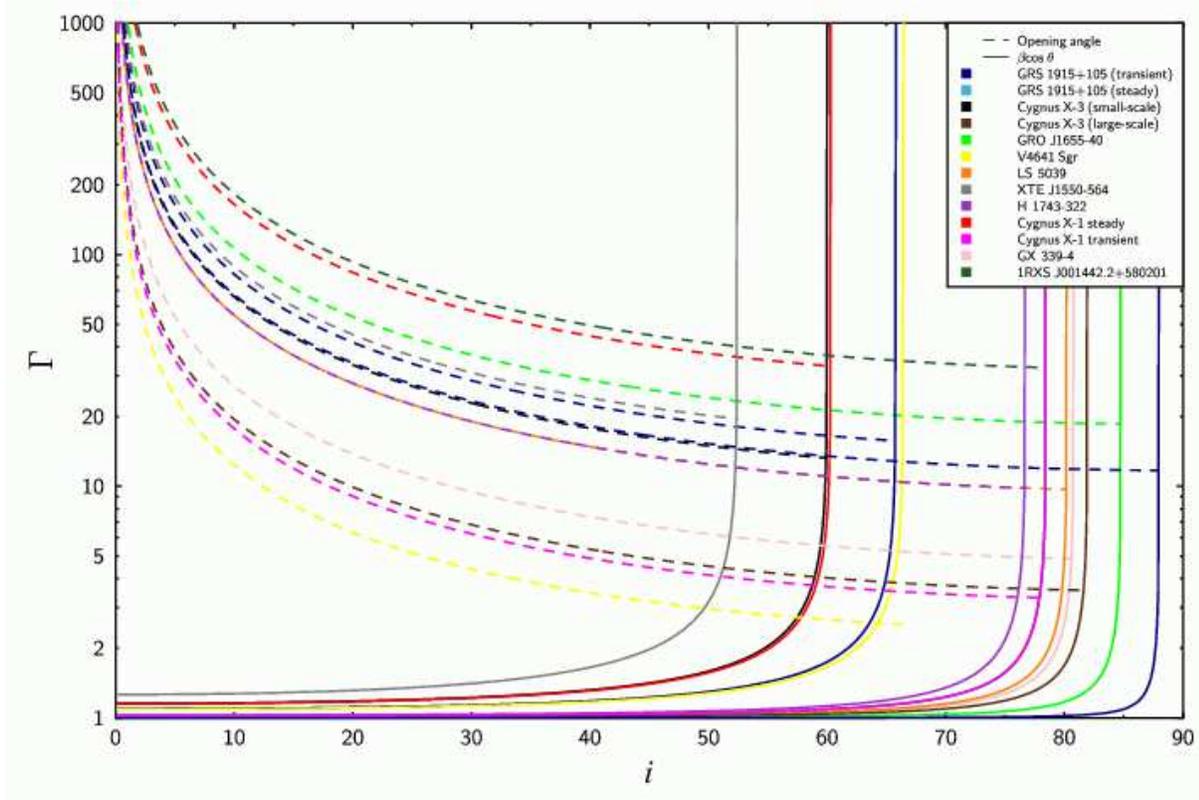}
\caption{Lorentz factors derived from opening angle considerations
  (Equation\,\ref{eq:gamma}; dashed lines) and from $\beta\cos i$
  (Table \ref{tab:obs}; solid lines) for a selection of Galactic XRBs.
  Since the opening angles given in Table \ref{tab:obs} are upper
  limits, the Lorentz factors derived from them (the dashed lines) are
  in fact lower limits, assuming freely-expanding jets with an
  expansion speed $c$.  The Lorentz factor $\Gamma_{\rm min,exp}$ is
  defined as the Lorentz factor where the dashed and solid lines
  cross, i.e.\ the solution for $\Gamma$ and $i$ satisfying both
  Equation\,\ref{eq:gamma} (for $\beta_{\rm exp}=1$ and $\phi$ as
  listed in Table \ref{tab:obs}) and the constraint on $\beta\cos i$.}
\label{fig:gamma}
\end{center}
\end{figure*}
Plots of the Lorentz factors calculated from Equation\,\ref{eq:gamma},
assuming free expansion at $c$ (i.e.\ $\beta_{\rm exp} = 1$) in the
source frame, are shown in Fig.\,\ref{fig:gamma} for the XRBs listed
in Appendix \ref{sec:details}, along with the Lorentz factors derived
from the constraints on $\beta\cos i$ given in Table \ref{tab:obs}.
Since Equation \ref{eq:gamma} contains four variables, then for
$\beta_{\rm exp}=1$ and $\phi$ fixed at the value in Table
\ref{tab:obs}, then with the constraint on $\beta\cos i$, there is a
unique solution for $\Gamma$ and $i$ for each source.  This is the
point at which the solid and dashed lines meet in
Fig.\,\ref{fig:gamma}.  We denote this value of $\Gamma$ as
$\Gamma_{\rm min,exp}$, which is listed in Table \ref{tab:obs}.  We
note that the plotted values of $\Gamma$ in Fig.\,\ref{fig:gamma} are
all strictly lower limits calculated for a freely-expanding source
($\beta_{\rm exp}=1$), since none of the jets (with the possible
exception of Cygnus X-3) were resolved perpendicular to the jet axis,
so we have only upper limits on the opening angles of the jets.  From
the figure, it is clear that the opening angles predict much higher
Lorentz factors (with a mean of 13.7) than are permitted by the
measurements of $\beta\cos i$, unless the source distances are all at
or very close to $d_{\rm max}$, the distance at which $i = i_{\rm
max}$ and $\beta=1$.  For the sources which are not very close to
$d_{\rm max}$ (see Table \ref{tab:obs} and \S\,\ref{sec:dmax}), then
the expansion speed must be less than $c$, i.e.\ {\it the jets are
confined}.

\begin{figure}
\begin{center}
\includegraphics[width=0.48\textwidth,angle=0,clip=]{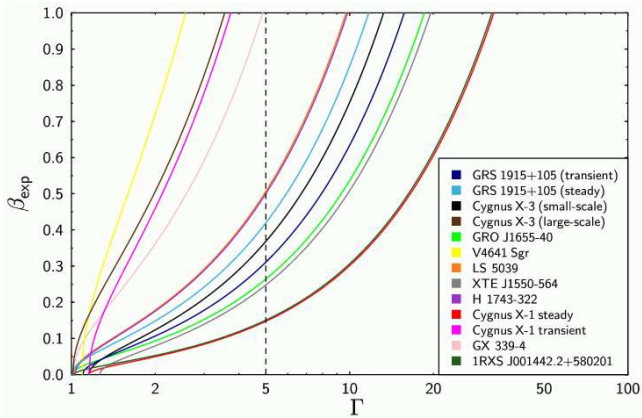}
\caption{Predicted expansion speeds for measured Lorentz factors from
Equation \ref{eq:beta_exp} and the constraints on $\beta\cos i$ and
$\phi$ given in Table \ref{tab:obs}.  The vertical dashed line
corresponds to a Lorentz factor of 5, showing that most jets must be
confined ($\beta_{\rm exp}<1$) if their observed opening angles are
determined by the transverse Doppler effect.}
\label{fig:beta_exp}
\end{center}
\end{figure}

If the jets are not freely expanding, but rather expand at some lower
velocity, $\beta_{\rm exp}c$, in their rest frame, then the jet Lorentz
factors implied by the measured opening angles are lower.  We can
rearrange Equation\,\ref{eq:gamma} as
\begin{equation}
\beta_{\rm exp} = \tan\phi\left\{\Gamma^2\left[1-(\beta\cos i)^2\right]-1\right\}^{1/2},
\label{eq:beta_exp}
\end{equation}
where, since $\beta > \beta\cos i$ as explained in Section
\ref{sec:literature}, then we require $\Gamma>[1-(\beta\cos
i)^2]^{-1/2}$.  Using the constraints on $\phi$ and $\beta\cos i$ in
Table \ref{tab:obs}, we have plotted the predicted values of
$\beta_{\rm exp}$ for different values of $\Gamma$ in
Fig.\,\ref{fig:beta_exp}.  This shows that only four of the jets have
Lorentz factors $\Gamma<5$ for $\beta_{\rm exp}=1$, as also seen in
Table \ref{tab:obs}.  If the Lorentz factors in XRB jets are all to
lie within the commonly-assumed range of $2<\Gamma<5$, then from
Fig.\,\ref{fig:beta_exp} we can quantify the degree of confinement
required.  In that case, the jets with the highest derived Lorentz
factors $\Gamma_{\rm min,exp}$ have a maximum possible expansion speed
of $\beta_{\rm exp}\lesssim 0.15$.  If the opening angles are indeed
smaller than the upper limits quoted in Table \ref{tab:obs}, then this
value would decrease further.

\subsection{Lorentz factors from jet power constraints}

The high bulk Lorentz factors derived from Equation\,\ref{eq:gamma}
would appear to imply high kinetic powers for the jets.  Nine of the
thirteen jets listed in Table\,\ref{tab:obs} have predicted
$\Gamma_{\rm min,exp}>9$.  Minimum energy requirements
\citep[e.g.][]{Lon94}, assuming that the source volume can be derived
by equating the light crossing time of the source to the rise time of
an outburst, give a minimum jet power of
\begin{multline}
P_{\rm min}(\Gamma=1) = 3.5\times10^{33}\eta^{4/7} \left(\frac{\Delta
    t}{{\rm s}}\right)^{2/7} \left(\frac{d}{{\rm
    kpc}}\right)^{8/7} \left(\frac{\nu}{{\rm
    GHz}}\right)^{2/7} \\
    \times \left(\frac{S_{\nu}}{{\rm
    mJy}}\right)^{4/7} \qquad \rm{erg}\,{\rm s}^{-1},
\end{multline}
where the source rise time $\Delta t$, the flux density $S_{\nu}$, and
the frequency $\nu$ should be measured in the source rest frame,
equivalent to the observer's frame for $\Gamma=1$.  $\eta$ is the
ratio of energy in relativistic electrons to the total energy.  Since
$\nu = \delta\nu^{\prime}$, $S_{\nu} =
\delta^{3-\alpha}S_{\nu}^{\prime}$ (for a spectrum
$S_{\nu}\propto\nu^{\alpha}$), and $\Delta t = \delta^{-1}\Delta
t^{\prime}$, then if $\Gamma\neq1$, the two frames are no longer
equivalent, and the correction
\begin{equation}
P_{\rm min}(\Gamma\neq 1) = \delta^{-4(3-\alpha)/7}P_{\rm min}(\Gamma=1),
\end{equation}
is required if the measured values in the observer's frame are to be
used.  In all the sources considered, $\delta<1$, such that increasing
$\Gamma$ implies an increase in the intrinsic source luminosity.

\citet{FBG04} tabulate measured values of $d$, $M/M_{\odot}$ (where
$M$ is the mass of the compact object), $\Delta t$, and $S_{\rm 5
GHz}$ for several sources, allowing us to calculate for each of the
sources the minimum power implied for the jets by the derived values
of $\Gamma_{\rm min,exp}$.  This can be compared with the Eddington
limiting luminosity, $L_{\rm Edd} =
1.3\times10^{38}(M/M_{\odot})$\,erg\,s$^{-1}$, shown as the ratio
$P_{\rm min}/L_{\rm Edd}$ in Table \ref{tab:lorentz}.  The values
shown for Cygnus X-3 have been taken from \citet{Mil04}, and those
shown for the 2001 flare of GRS\,1915+105 ($\phi =
18.3\pm3.6^{\circ}$) from \citet{Mil05}, to give an illustration of
how variable flare power can be.

\begin{table*}
\begin{center}
\begin{tabular}{lccccccc} \hline \hline
System & $d$ (kpc) & $M/M_{\odot}$ & $\Delta t$ (s) & $S_{\rm 5
  GHz}$ (mJy) & $\beta\cos i$ & $P_{\rm min}/L_{\rm Edd}$ & $\Gamma_{\rm
  min,exp}$\\
\hline
GRS\,1915+105 (2001 transient) & 11 & 14 & 21000 & 100 & $0.29\pm0.09$ &
  0.03 & 3.3\\
GRS\,1915+105 (1997 transient) & 11 & 14 & 43200 & 320 & $0.41\pm0.02$ &
  1.87 & 15.7\\
GRS\,1915+105 (steady) & 11 & 14 & 300 & 50 & $0.035\pm0.017$ &
  0.09 & 11.7\\
Cygnus X-3 (small-scale) & 10 & $\sim 4$? & $3\times 10^5$ & 13400 & $0.50\pm0.10$ & 43.3 & 13.3\\
GRO\,J1655--40 & 3.5 & 7 & 43200 & 2000 & $0.091\pm0.014$ & 13.9 & 18.6\\
V4641\,Sgr & 8 & 9 & 43200 & 420 & $\sim 0.4$ & 0.22 & 2.6\\
XTE\,J1550--564 & 6 & 9 & 43200 & 130 & $0.61\pm0.13$ & 1.92 & 19.6\\
Cygnus X-1 (steady) & 2.5 & 10 & 2000 & 50 & $>0.50$ & 0.30 & 33.0\\
GX\,339--4 & 8 & 7 & 19800 & 55 & $>0.16\pm0.05$ & 0.24 & 4.9\\
\hline \hline
\end{tabular}
\caption{Measured parameters needed to calculate jet power $P_{\rm
    min}$ from minimum energy arguments.  Data taken mainly from
    \citet{FBG04}.  The Lorentz factors derived from opening angle
    considerations are not ruled out by jet power constraints except
    in the case of Cygnus X-3.}
\label{tab:lorentz}
\end{center}
\end{table*}

It is plausible that XRBs can exceed the Eddington limiting luminosity
for a short time during outburst by a small factor.  Thus the Lorentz
factors implied by the measured opening angle constraints are clearly
ruled out by the total power requirement only in the case of Cygnus
X-3, a source which we believe for other reasons to be confined
(\S\,\ref{sec:known_confined}).  Thus in most cases, jet power
constraints cannot rule out high Lorentz factors.

\subsection{Lorentz factors from source distances}
\label{sec:dmax}
By definition, all significantly relativistic jets should lie close to
$d_{\rm max}$ (as defined in Equation \ref{eq:dmax}).  Therefore, for
sources at distances significantly less than $d_{\rm max}$, we can
constrain their Lorentz factors via
\begin{equation}
\Gamma = \left\{1-\left(\frac{d}{d_{\rm max}}\right)^2\left[1-(\beta\cos
i)^2\right]-(\beta\cos i)^2\right\}^{-1/2}.
\label{eq:gamma_d}
\end{equation}
The derived values, $\Gamma_{\rm d}$, are listed in Table
\ref{tab:obs} for the four sources with measured proper motions (such
that $d_{\rm max}$ could be calculated) and independently-determined
source distances.  Since $\Gamma$ increases rapidly as $d$ approaches
$d_{\rm max}$, we can only put a lower limit on the Lorentz factors of
the two sources lying close to (within $2\sigma$ of) $d_{\rm max}$,
GRS\,1915+105 and GRO J\,1655--40.  On the other hand, Cygnus X-3 and
XTE\,J1550--564 both appear to lie at $d<d_{\rm max}$, so neither
would appear to have a significantly relativistic jet.

\subsection{Jets known to be confined}
\label{sec:known_confined}
Further evidence for the jets in Cygnus X-3 and XTE\,J1550--564 being
confined comes from the observed deceleration of the jets.  In the
milliarcsecond-scale jets of Cygnus X-3 \citep{Mio01,Mil04}, the
southern jet was found to be approaching and the northern counterjet
receding from us.  On arcsecond scales however, the northern component
is both brighter and at a greater angular separation than the southern
component \citep{Mar01}, suggesting that the southern component has
decelerated on moving outwards.  It must therefore have a low bulk
Lorentz factor, $\lesssim 2$, on arcsecond scales at least.  The
distance discrepancy addressed in \S\,\ref{sec:dmax} suggests that the
bulk Lorentz factor is also low on milliarcsecond scales.  The X-ray
jets of XTE\,J1550--564 showed a measurable decrease in the rate of
angular separation from the core with time \citep{Kaa03}.  In order to
produce a measurable deceleration in the proper motions, the bulk
Lorentz factor of the component must be $\lesssim 2$, since
significant changes in proper motions for a (presumably) fixed angle
to the line of sight are only possible when $\beta$ is changing
appreciably, i.e.\ in the regime $\Gamma\lesssim 2$.  It is possible
that the X-ray jets detected in H\,1743--322 were also similarly
powered by bulk deceleration, although the observations cannot confirm
this.  For Cygnus X-3 and XTE\,J1550-564 therefore, since
$\Gamma\lesssim 2$, Fig.\,\ref{fig:beta_exp} shows that the jets must
be confined.

Although we have not plotted the Lorentz factor for SS\,433 in
Fig.\,\ref{fig:gamma}, the jets in this source are certainly confined.
The opening angles inferred from observations are $<7$\degree
(Appendix \ref{sec:details}), whereas the predicted jet opening angle
from Equation~\ref{eq:opening_angle} is at least 74\degree for
transverse expansion at $c$, given the known bulk velocity of $0.26c$
\citep{Abe79,Hje81}.  Confinement (on small scales at least) was
proposed for this source by \citet{Hje88}, who suggested that the jet
underwent a transition from slowed to free expansion (i.e.\ became
unconfined) at a distance of $\sim25$ light days from the core.  We
know that the jets of SS\,433 contain baryons, so it is possible that
these cold protons could be responsible for the slowed expansion (see
\S\,\ref{sec:confinement}).

\section{Low-hard state jets}
\label{sec:low-hard}
Two different manifestations of jets are known to exist in XRBs;
steady, flat-spectrum outflows observed in the low/hard X-ray state
\citep[e.g.][]{Dha00b}, and discrete superluminal ejecta seen during
transient outbursts \citep[e.g.][]{Mir94}.  According to the internal
shock model of Fender, Belloni \& Gallo (2004a), the Lorentz factors
of the steady jets should be lower than for the transient jets.  Our
sample contains both steady and transient jets in Cygnus X-1 and
GRS\,1915+105, so we can compare the derived Lorentz factors to see if
there is a difference.

Steady jets are unlikely to be able to exceed the Eddington limit, a
condition satisfied by the values of $P_{\rm min}/L_{\rm Edd}$ given
in Table \ref{tab:lorentz}.  Comparing the Lorentz factors derived
from opening angle considerations, $\Gamma_{\rm min,exp}$ (listed in
Table \ref{tab:obs}), for the transient and steady jets, there is no
obvious trend.  $\Gamma_{\rm min,exp}$ for the steady jet in
GRS\,1915+105 is greater than that derived for the 2001 flare, but
less than that for the 1997 flare.  However, since in neither Cygnus
X-1 nor in GRS\,1915+105 were the jets resolved, the opening angles we
have listed are all upper limits, and thus we would not necessarily
expect to see a significant difference between the steady and the
transient jets.

We cannot constrain the steady jets to be less relativistic than the
transient jets from their opening angles.  Therefore, if they are
indeed significantly less relativistic, as predicted by the internal
shock model, then the steady jets would have to be confined.

\section{Comparison to AGN}
\label{sec:AGN}
XRB jets are in general thought to be less relativistic than those in
AGN.  Their typical bulk Lorentz factors are commonly assumed to be of
order 2--5, as compared to the AGN jets with bulk Lorentz factors up
to $\sim 20$.  A recent discovery of an ultrarelativistic outflow at
$\beta_{\rm app}>15$ in Circinus X-1 \citep{Fen04}, later revised down
to $\beta_{\rm app}>9.2$ by \citet{Iar05}, challenges the assumption
that stellar-mass objects cannot produce highly-relativistic jets.
From Equation \ref{eq:vel_apparent}, the apparent component velocity
$\beta_{\rm app}$ has a minimum value of $\Gamma\beta$ when $\beta =
\cos i$, where $\beta$ is the true component velocity.  $\beta_{\rm
app}$ is thus a lower limit to the value of $\Gamma$.  Hence XRBs are
clearly capable of producing jets with Lorentz factors $\sim 10$.

\begin{figure}
\begin{center}
\includegraphics[width=0.45\textwidth,angle=0,clip=]{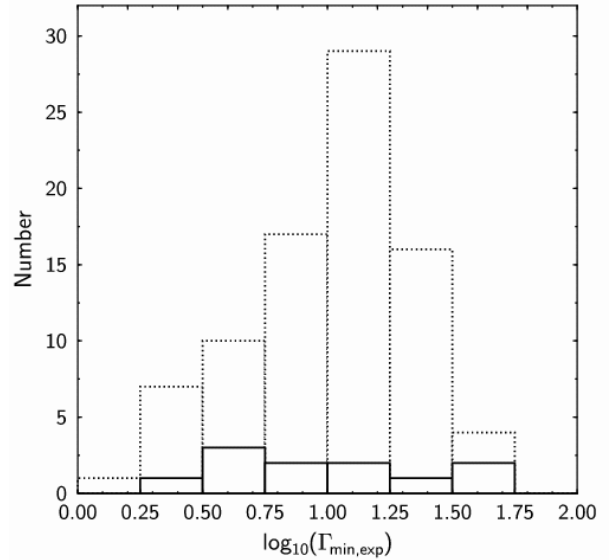}
\caption{Distributions of Lorentz factors derived for XRB
  jets and AGN jets.  The solid bars show the distribution for X-ray
  binary jets, assuming that they all have the minimum Lorentz factor
  permitted from opening angle considerations, and assuming free jet
  expansion.  The dashed bars show the distribution for AGN jets,
  calculated using the proper motions tabulated by \citet{Jor01},
  assuming that they all have the minimum permitted bulk Lorentz
  factors, with $\beta = \cos i$.}
\label{fig:gamma_distributions}
\end{center}
\end{figure}

Proper motions for jet components in a sample of $\gamma$-ray bright
blazars were measured by \citet{Jor01}.
Fig.\,\ref{fig:gamma_distributions} shows the minimum bulk Lorentz
factors derived from their tabulated values of $\beta_{\rm app}$
compared to those derived from the opening angle constraints for XRB
jets in \S\,\ref{sec:opening_angle}.  The Lorentz factors for jets
found to be confined in \S\,\ref{sec:known_confined} have been omitted
from the histogram.  Since the sample is so small for the X-ray
binaries, and since in both cases we have plotted lower limits on the
Lorentz factors, a quantitative comparison is not possible.  However,
the histograms do show that {\it if} XRB jets are unconfined, and the
derived opening angles are purely due to the relativistic effect
described in \S\,\ref{sec:formalism}, then XRB jets can be at least as
relativistic as AGN jets.

\subsection{High-mass X-ray binaries}
While Circinus X-1 is known to produce jets with high bulk Lorentz
factors, no high-mass XRB is known to produce highly relativistic jets
able to move ballistically outwards at $\beta\sim 1$ with no
deceleration.  We have already discussed the cases of SS\,433 and
Cygnus X-3.  The transient jets of Cygnus X-1 were thought to have
velocities $v\gtrsim0.5c$, estimated from an assumed ejection date
(Fender et al., in prep.).  In CI Cam, no jet was seen during the 1998
outburst, but rather an expanding radio nebula, consistent with a
shock moving through a dense stellar wind \citep{Mio04}.  We therefore
suggest that it is only the low-mass XRBs that can produce resolved
jets with the high bulk Lorentz factors discussed herein, as a result
of the more tenuous ambient medium.  This does not mean that the
high-mass XRBs are not launched with comparable Lorentz factors, but
rather that their stronger interactions with the surrounding ISM tend
to decelerate the jets before they become resolved.

\section{Confinement mechanisms}
\label{sec:confinement}

Both Begelman, Blandford \& Rees (1984) and \citet{Fer98} give
detailed discussions on possible jet confinement mechanisms.  The most
natural confining agent for a jet is thermal gas pressure from an
external medium or a magnetic field.  The internal jet pressure,
$p_{\rm min}$, can be calculated from equipartition arguments, since
\begin{equation}
p_{\rm min} = (\gamma_{\rm SH}-1)U_{\rm min} = (\gamma_{\rm
  SH}-1)\frac{E_{\rm min}}{V},
\end{equation}
where $U_{\rm min}$ is the minimum energy density in the source, and
$\gamma_{\rm SH}$ is the ratio of specific heats, equal to 4/3 for an
ultrarelativistic gas.  A minimum energy density can be found by
dividing the minimum energy $E_{\rm min}$ corresponding to the
observed synchrotron emission by the source volume, $V$.  If the
pressure of the external medium is given by the ideal gas law,
$p=nk_{\rm B}T$, where $k_{\rm B}$ is Boltzmann's constant, $n$ the
number density of particles, and $T$ the temperature, then typical ISM
number densities and temperatures give pressures several orders of
magnitude too low to confine the jets.  This is unsurprising, since
\citet{Hei02} found that microquasars inhabit low-density,
low-pressure environments when compared in a dynamical sense to the
environments of AGN.  We do not therefore consider pressure
confinement by the ambient ISM to be a possibility.

One of the major differences between microquasars and AGN is that AGN
jets tend to be surrounded by observable cocoons of waste plasma
flowing back from the hotspots at the ends of the jets.  This plasma
is at higher pressure than the ambient ISM, helping to confine the
jet.  Such cocoons have only been detected in a few Galactic systems.
Cygnus X-1 is known to have inflated a bubble in the interstellar
medium, whose bowshock has been imaged \citep{Gal05}, although the
surface brightness of the lobes themselves was too low to be detected.
The jets in SS\,433 have deformed the supernova remnant W\,50,
creating a pair of lateral extensions with position angles in perfect
agreement with the jet axis \citep[e.g.][]{Dub98}.  Possible
hotspot-like structures have recently been tentatively associated with
the sources GRS\,1915+105 \citep{Rod98,Cha01,Kai04} and Cygnus X-3
\citep{Mar05}, although any association with the source is yet to be
definitively confirmed in either case.  Lobe structures have, as
stated in Appendix \ref{sec:details}, been detected in the sources
1E1740.7--2942 and GRS\,1758--258, but in neither source have the jets
themselves been observed.  \citet{Hei02} found that since microquasar
sources stay relativistic for a dynamically longer time than AGN, and
are located in underdense environments, their detectability will be
severely limited and only possible at low frequencies.  Such
low-frequency observations with new and upcoming facilities such as
the new Low Frequency Front Ends on the Westerbork Synthesis Radio
Telescope, and LOFAR \citep[the Low Frequency Array;][]{Rot03} may
shed light on this issue.

\citet{Ick92} suggested that jets could be inertially confined (on
small scales at least) by an outflowing wind around the jet, whose ram
pressure would oppose the jet expansion.  \citet{Pet95} investigated
inertial confinement, and found that inertial effects could collimate
a jet provided that the material to be collimated was sufficiently
tenuous compared to the surrounding flow responsible for the
collimation.

A toroidal magnetic field can in theory confine a jet via the hoop
stress mechanism, although toroidal fields are notoriously unstable to
the kink instability, which could disrupt the jet.  Furthermore,
\citet{Eic93} found that magnetically confined jets are subject to
only a modest amount of collimation in the absence of an additional
collimating mechanism.  So far from the anchor point of the magnetic
field, some dynamo effect would likely be required to maintain a
sufficiently strong magnetic field.  However, Spruit, Foglizzo \&
Stehle (1997) suggested that poloidal fields could lead to collimation
at some collimation distance of order the disc radius, after which the
jet would remain collimated, expanding ballistically with a fixed
opening angle out to large distances.  If the jet material was simply
streaming ballistically outwards, no collimating mechanism would be
required.

A final possibility is that the jet could contain cold material;
either protons or pairs.  Since we observe synchrotron radiation in
the radio band, we know that there are highly relativistic electrons
present in the jets.  This could just be the high-energy tail of a
thermal population of pairs, although we consider this unlikely given
the observed relativistic bulk velocities.  However, if cold protons
(nonrelativistic in the comoving frame) were present, they would
dominate the inertia of the jets unless the mean electron Lorentz
factors were $\gtrsim 2000$.  Applying minimum energy arguments to the
X-ray emitting knots in XTE\,J1550--564 \citep{Tom03} gives a mean
electron Lorentz factor of $7.8\times10^3<\langle\gamma_{\rm
e}\rangle<2.9\times10^4$, depending on the source distance.  In this
case, assuming equipartition of energy between protons and electrons,
even the protons would have $\langle\gamma_{\rm p}\rangle>4.25$.  So
cold protons do not in all cases dominate the inertia of XRB jets, but
if they were to do so, the expansion would be retarded compared to a
plasma consisting solely of electron-positron pairs.  This confinement
mechanism has been effectively ruled out in AGN by \citet{Cel98}, who
found that the filling factor of the jet with thermal material had to
be so small as to be insignificant in terms of the overall jet energy
budget.  As mentioned in \S\,\ref{sec:known_confined},
this mechanism could be responsible for confinement in SS\,433.

\subsection{Lightcurves: a possible test of confinement}
\label{sec:lc}
A possible test of whether the jets are confined could be made
using jet lightcurves.  The magnetic field in the jet scales as
$V^{-2/3}$ for a non-turbulent jet, where $V$ is the jet volume.  The
Lorentz factor of the particles in an adiabatically expanding jet
scales as $V^{-1/3}$, and thus the jet emissivity scales as $J\propto
V^{(4\alpha-2)/3}$.  For a jet expanding freely at constant velocity,
$V\propto t^3$, so the flux density scales as $S\propto
t^{(4\alpha-2)}$.  The magnetic field would fall off more slowly with
jet volume for a turbulent jet (typically as $V^{(-2+\zeta)/3}$, where
$\zeta\sim 1$ describes the degree of turbulence).  We note however
that once the magnetic field energy has reached equipartition with the
internal energy, a decay slower than $B\propto V^{-1/2}$ is not
possible on energetics grounds.  Fitting the jet lightcurves in the
optically-thin part of the spectrum with a power-law decay,
$S=S_0(t/t_0)^{-\xi}$, we can then place limits on the confinement of
the jets.  For a typical optically-thin jet, $\alpha\sim-0.6$, and
since the power law index $\xi$ is Lorentz invariant, we would expect
$\xi\sim 4$ if the jets are freely expanding (the exact value
depending on the degree of turbulence).  If $\xi$ is much less than
this ($\lesssim 3$), the jets are not expanding at constant velocity,
which strongly indicates jet confinement, as there is no reason that
the volume of a confined jet would scale as $t^3$.

To measure the flux density decay with time, the jet lightcurve must
be decoupled from that of the core, which requires both resolving the
jets, and measuring their flux density at more than two epochs.
Nevertheless, \citet{Rod99} measured power-law indices of $1.3\pm0.2$
close to the core and $2.6\pm0.5$ at distances $>2\times10^{17}$\,cm
in GRS\,1915+105, suggesting a switch from confined to free expansion
(as previously suggested for SS\,433).  \citet{Mil05} measured
$1.80\pm0.03$ and $2.01\pm0.02$ in the same source for two outbursts
in 2001 on scales smaller still, confirming the confinement at small
angular separations.  The X-ray flux in XTE\,J1550--564, on the other
hand, was observed to decay with a power-law index of $3.7\pm0.7$
\citep{Kaa03}, so the X-ray jets in this source could be expanding at
constant velocity.  Since we found in \S\,\ref{sec:known_confined}
that the jets in this source are almost certainly confined, the
expansion velocity is likely to be less than $c$, if constant.

In summary, we cannot easily rule out many of the possible confinement
mechanisms in XRB jets, and the evidence from the decay lightcurves of
the two sources considered above suggests that the jets are not
freely-expanding.  Should the jets be confined, they would not need to
be as relativistic as implied by the opening angle calculations of
\S\,\ref{sec:opening_angle}.  But given the differences in conditions
between the environments of Galactic and extragalactic objects
outlined in this section, and the likely variations in those
properties within the Galaxy, it is at least plausible that XRB jets
need not all be confined.

\section{Conclusions}
If XRB jets are not confined, but are expanding freely, it is possible
to constrain their Lorentz factors from measurements of the jet
opening angles.  The small opening angles we observe are in this case
a consequence of the transverse Doppler effect slowing the apparent
expansion speed in the observer's frame. From the upper limits to the
opening angles quoted in the literature, the Lorentz factors thus
derived are significantly more relativistic (with a mean Lorentz
factor $>10$) than is commonly assumed, and possibly no less so than
AGN jets.  However, if the jets we observe do indeed have Lorentz
factors in the commonly-assumed range of 2--5, then we can quantify
the degree of confinement.  The lateral expansion speed perpendicular
to the jet axis must then in some cases be $\lesssim 0.15c$.

In most cases, we cannot exclude the possibility that the jets are
unconfined from jet power constraints, nor from measurements of the
proper motions of knots in XRB jets.  From the distances $d<<d_{\rm
max}$ and the observed deceleration in the jets of Cygnus X-3 and
XTE\,J1550--564, we know that the jets in these sources at least must
be confined.  The observed opening angle of the jets in SS\,433
suggests that its jets are also confined, possibly by the cold protons
known to exist in the jets.  However, we cannot definitively rule out
confinement in any of the other sources considered.  We do not find
any evidence for a difference in the velocities of low/hard state jets
and transient jets, although the observations only provided lower
limits to the Lorentz factors in both cases.

In many cases, and as observed in Circinus X-1, XRB jets could well be
significantly more relativistic than is commonly assumed, although
this is unlikely in the case of high-mass XRBs.  While we cannot rule
out many of the possible confinement mechanisms, in the absence of
definitive lower limits to the jet Lorentz factors, the possibility
that XRB jets are highly relativistic ($\Gamma \sim 10$) should not be
ruled out.

\section*{Acknowledgments}
We would like to thank Christian Kaiser, Asaf Pe'er and Vivek Dhawan
for useful discussions.  JM-J thanks the University of Southampton and
the Leids Kerkhoven-Bosscha Fond for support during his visit to
Southampton.

\label{lastpage}

\bibliographystyle{mn2e}

\appendix
\section{The individual sources}
\label{sec:details}
Here we present a summary of the observations on the individual X-ray
binaries which have been observed to exhibit resolved jets.

\noindent
\textit{GRS\,1915+105:} Discrete knots in this system have been followed
as they moved outwards from the core \citep{Mir94,Fen99,Mil05}.  We
chose the highest-resolution observations \citep{Fen99} to constrain
$\beta\cos i = 0.41\pm0.02$.  From the lack of resolved structure
perpendicular to the jet axis, the half-opening angle is constrained
to be $\leq 4$\degree.  Milliarcsecond-scale radio jets have also been
imaged during the X-ray hard plateau state in this source (Dhawan,
Mirabel \& Rodr\'\i guez 2000b; Fuchs et al.\ 2003; Rib\'o, Dhawan \&
Mirabel 2004).  \citeauthor{Dha00b} found that these jets were
marginally resolved perpendicular to the jet axis at 43\,GHz, although
in this case the opening angle cannot be derived as the black hole
position is not known (V.\ Dhawan, private communication).  From the
beamsize and jet length in their lower-resolution 15-GHz images, we
derived a constraint on the half-opening angle of $\leq 5$\degree.

\noindent
\textit{Cygnus X-3:}
Again, discrete knots have been followed as they moved out from the
core of the system \citep{Mil04}.  The ratio of proper motions of
approaching and receding jet components constrained $\beta\cos i =
0.50\pm0.10$, and the measured knot size perpendicular to the jet axis
from the last high-resolution 22-GHz image was used to constrain the
half-opening angle to be $5.0\pm0.5$\degree.  On larger scales,
Mart\'\i, Paredes \& Peracaula (2001) found $\beta\cos i>0.14\pm0.03$
and did not resolve the knots perpendicular to the jet axis at an
angular separation of $\sim0.6$\,arcsec with a beam diameter of
361\,mas, giving an upper limit on the half-opening angle of
$\phi<16.5$\degree.

\noindent
\textit{GRO\,J1655--40:} 
The highest-resolution observations of the 1994 outburst of
GRO\,J1655--40 were made by \citet{Hje95} at 1.6\,GHz with the VLBA.
The ejecta were observed moving out to a maximum separation of
$\sim1^{\prime\prime}$, and the beamwidth was $\sim43\times108$\,mas,
with no evidence for the jets being resolved.  This constrained
the opening angle to $\leq6.1$\degree.  From the ratio of proper
motions of the NW and SE ejecta, $\beta\cos i = 0.091\pm0.014$, and
the kinematic model gives a distance of $3.2\pm0.2$\,kpc.

\noindent
\textit{V4641 Sgr:}
\citet{Hje00} observed the 1999 September outburst of V4641 Sgr at
4.9\,GHz with the VLA, and found an extended jetlike structure at an
angular separation of 0.25\arcsec.  The restoring beam was a circular
Gaussian of FWHM 0.3\arcsec, constraining the opening angle to be
$\leq 50.2$\degree.  From the kink in the lightcurve of the outburst,
attributed to the offset between the peaks in the approaching and
receding jets, they found $\beta\cos i=0.4$.  They argued that the
source distance lies in the range $0.4\leq d \leq 1.7$\,kpc with the
nearer distance being most likely, in stark contradiction to the
distance of $9.59^{+2.72}_{-2.19}$\,kpc found from the luminosity of
the secondary star \citep{Oro01}.

\noindent
\textit{LS\,5039:}
A two-sided milliarcsecond-scale radio jet in LS\,5039 was observed at
5\,GHz with the EVN and MERLIN \citep{Par02}.  Using the length
asymmetry of the jets on either side of the core, $\beta\cos i =
0.17\pm0.05$ and since the jet width was smaller than one synthesised
beam, the half-opening angle was constrained to $\leq 6$\degree.

\noindent
\textit{XTE\,J1550--564:} \citet{Cor02} first detected radio and X-ray
emitting jets from XTE\,J1550--564, which they found to be
decelerating with time.  \citet{Tom03} found no evidence for extension
perpendicular to the jet axis in the approaching (eastern) jet,
constraining the opening angle component to be $<7.5$\degree\ at an
angular separation of $23.4\pm0.5$\arcsec.  In observations taken 2
years later, \citet{Kaa03} found weak evidence for extension
perpendicular to the jet axis in the receding (western) X-ray jet at a
similar angular separation, giving a more stringent limit on the half
opening angle of $<1$\degree.  From the measured proper motions of the
eastern and western jets, assuming symmetric jet propagation, they
found $\beta\cos i=0.61\pm0.13$ and $d_{\rm max}=16.5\pm3.5$\,kpc.
The true distance was found to be in the range $1.4<d<9.8$\,kpc from
the observed luminosity of the secondary star, and $3.2<d<10.8$\,kpc
from the systemic velocity, with a favoured value of 5.3\,kpc
\citep{Oro02}.  For consistency with other sources, we will use the
half-opening angle constraint of \citet{Tom03}, since it is measured
for the approaching jet.

\noindent
\textit{H\,1743--322:} 
\citet{Cor05} detected moving X-ray jets from another microquasar
system, H\,1743--322.  Assuming ballistic jet motion, they found
$\beta\cos i = 0.23\pm0.05$, and $d_{\rm max}=10.4\pm2.9$\,kpc.  The
upper limit on the source FWHM was found to be 1.4\arcsec\ at an
angular separation of 6.63\arcsec, implying an opening angle of
$<12$\degree.  The source lies in the direction of the Galactic Centre,
so its distance was assumed to be 8.5\,kpc, although no definitive
measurement has yet been made.

\noindent
\textit{Cygnus X-1:}
\citet{Sti01} imaged an extended jet-like radio structure in Cygnus
X-1 while the source was in the low/hard X-ray state.  The jet was not
resolved perpendicular to the flow, constraining the half-opening angle to
be $\leq 2$\degree.  From the ratio of the flux density in the
detected jet to the upper limit on the flux density in the receding
jet (the r.m.s.\ noise in the image multiplied by the area occupied by
the approaching jet), a value for $\beta\cos i$ may be obtained from
the equation
\begin{equation}
\frac{S_{\rm app}}{S_{\rm rec}} = \left(\frac{1+\beta\cos
  i}{1-\beta\cos i}\right)^{k-\alpha},
\label{eq:ratio}
\end{equation} 
where $k=2$ for a continuous jet and $k=3$ for a jet composed of
discrete knots.  Taking $k=3$ and a spectral index $\alpha=-0.6$ gives
the minimum likely value $\beta\cos i>0.5$.  More recently, Fender et
al.\ (in prep.) have detected a resolved transient jet from this
source at an angular separation of $70\pm5$\,mas, unresolved with a
beamsize of 50\,mas, implying an opening angle of 36\degree.  The bulk
jet velocity was found to be $\gtrsim0.5c$, which, if the jet and disc
are aligned ($i=30$\degree) implies $\beta\cos i>0.2$.

\noindent
\textit{GX\,339--4:} 
After an outburst in 2002 May, \citet{Gal04}
detected variable extended structure in GX\,339--4.  At the latest
epoch, the knot was separated from the core by 6.9\arcsec with a
beamwidth of 3.06\arcsec, implying an opening angle of $\leq
24$\degree.  From Equation\,\ref{eq:ratio} and the known spectral
index of $-0.98\pm0.10$ gives a constraint $\beta\cos i \geq 0.56$.
The source distance is an issue of considerable debate.  Systemic
velocity measurements gave $4\pm1$\,kpc \citep{Zdz98}, whereas the
upper limit on the magnitude of the secondary star implies
$d>5.6$\,kpc (Shahbaz, Fender \& Charles 2001).  \citet{Mac03}
estimated $d>7.6$\,kpc from the X-ray state transition luminosity, and
interstellar absorption measurements suggest a distance of at least
6\,kpc, and possibly as high as 15\,kpc \citep{Hyn04}.  The distance
should therefore be considered highly uncertain.

\noindent
\textit{1RXS\,J001442.2+580201:}
Identified as a microquasar candidate by Paredes, Rib\'o \& Mart\'\i\
(2002b), resolved jets were detected in this source by \citet{Rib02b}.
The latter also found a 3-sigma result for detectable source proper
motion suggestive of a Galactic nature, for which reason this source
is included in our sample.  From the angular separations of the
components from the core, they found $\beta\cos i = 0.20\pm0.02$.  The
components were unresolved with the EVN beam of 0.86\,mas, giving a
constraint on the opening angle of $<3.6$\degree.  With no detected
component proper motions, no constraint can be placed on $d_{\rm
max}$.

\noindent
\textit{SS\,433:} 
The most stringent (albeit model-dependent) limits on the opening
angle of the jet in SS\,433 come from the widths of X-ray lines
measured by Marshall, Canizares \& Schulz (2002), ascribed to the
Doppler broadening due to a conical outflow of constant opening angle.
The half opening angle of the jet was found to be
$0.61\pm0.03$\degree, which may be compared to the upper limit of
6.8\degree\ found from radio observations \citep{Ver93}, and to that
of $\lesssim 5$\degree\ found from the widths of the optical emission
lines \citep{Beg80}.  The jet velocity is known to be $\beta =
0.26\pm0.05$ and the inclination angle of the jet axis to the line of
sight is 80\degree\ \citep{Hje81}, although since the jet precesses,
the inclination angle to the line of sight changes, so we have not
included this source in our calculations.

The two confirmed neutron stars with resolved jets, Scorpius X-1
(Fomalont, Geldzahler \& Bradshaw 2001) and Circinus X-1,
\citep{Fen04} have not been included in this survey, since it seems
that the lobes we see are the working surface where an
ultrarelativistic, unseen flow of energy impacts on the ambient
medium.  It is therefore not possible to constrain the opening angle
of the jet from the sizes of the lobes.  Also, radio lobes have been
detected in the black hole candidates 1E\,1740.7--2942 \citep{Mir92}
and GRS\,1758--258 (Rodr\'\i guez, Mirabel \& Mart\'\i\ 1992;
Mart\'\i\ et al.\ 2002), but the lack of proper motion in both sources
again argues for these lobes being hotspots where the jets impact the
ISM, so they cannot be used to constrain the jet opening angles.

\end{document}